\documentstyle[11pt,newpasp,twoside,epsf]{article}
\reversemarginpar

\begin{document}
\title{Millisecond Pulsars in 47 Tucanae}
\author{Paulo C. Freire$^1$, Fernando Camilo$^1$, Duncan R.
Lorimer$^2$, Andrew G. Lyne$^1$, Richard N. Manchester$^3$}
\affil{$^1$U. of Manchester, Jodrell Bank Observatory, Cheshire,
SK11~9DL, UK}
\affil{$^2$NAIC, Arecibo Observatory, HC3 Box 53995, Arecibo, PR~00613, USA}
\affil{$^3$ATNF, CSIRO, P.O.~Box~76, Epping NSW~1710, Australia}

\begin{abstract}
Recent observations of the globular cluster 47~Tuc, made with the
Parkes telescope at a wavelength of 20\,cm, have resulted in the
discovery of nine new millisecond pulsars (MSPs), all in binary
systems.  The number of timing solutions available has risen from two
to 14.  These results will make possible a more detailed study of the
cluster dynamics.
\end{abstract}

\begin{tiny}

\begin{table}
\begin{center}
\caption{The 20 pulsars known in 47~Tuc. The solutions are based on
data taken between August 1997 and July 1999.  $R_{c}$ is the pulsar's
projected distance to the cluster centre in units of $r_{c}$, the core radius
($12.2''$).  Asterisks denote binary pulsars.}
\begin{tabular}{c l l l l}
\tableline
Pulsar & Period & $\dot{P}$           & Dispersion measure & $R_{c}$   \\
       & (ms)   & ($\times 10^{-20}$) & ($\rm cm^{-3} pc$) & ($r_{c}$) \\
\tableline
\multicolumn{5}{c}{Previously known pulsars} \\
\tableline
C & 5.75678      & $-$4.99(1)   & 24.60(1)  &  5.9  \\
D & 5.35757      & $-$0.28(1)   & 24.72(2)  &  3.4  \\
E & 3.53633$^*$  & \,~~9.860(8) & 24.24(2)  &  3.3  \\
F & 2.62358      & \,~~6.459(6) & 24.39(2)  &  0.8  \\
G & 4.04039      & $-$4.21(3)   & 24.5(1)  &  1.4  \\
H & 3.21034$^*$  & $-$0.17(1)   & 24.4(1) &  3.7  \\
I & 3.48499$^*$  & $-$4.58(3)   & 24.5(1) &  1.4  \\
J & 2.10063$^*$  & $-$0.976(1)  & 24.59(2)  &  4.8  \\
L & 4.34618      &$-$12.23(2)   & 24.4(1) &  0.7  \\
M & 3.67664      & $-$3.83(5)   & 24.4(2)   &  5.2  \\
N & 3.05395      & $-$2.18(9)   & 24.57(3)  &  2.4  \\
\tableline
\multicolumn{5}{c}{Newly discovered pulsars} \\
\tableline
O & 2.64334$^*$  &  ~~2.99(2)   & 24.4(1)  &  0.3  \\
P & 3.64302$^*$  &  ~~\dots     & 24.4(2)  & \dots \\
Q & 4.03318$^*$  &  ~~3.43(3)   & 24.3(1)  &  4.8  \\
R & 3.48046$^*$  &  ~~\dots     & 24.3(2)  & \dots \\
S & 2.830$^*$    &  ~~\dots     & 24.3(2)  & \dots \\
T & 7.589$^*$    &  ~~\dots     & 24.4(1)  & \dots \\
U & 4.34283$^*$  &  ~~9.47(4)   & 24.3(1)  &  4.6  \\
V & 4.81$^*$     &  ~~\dots     & 24.2(1)  & \dots \\
W & 2.35234$^*$  &  ~~\dots     & 24.4(2)  & \dots \\
\tableline
\tableline
\end{tabular}
\end{center}
\end{table}

\end{tiny}

\section{Pulsar searching}

The cluster 47~Tucanae (47~Tuc) was long known to contain 11 MSPs, with
four of them in binary systems (Robinson et~al.~1995).  In 20\,cm
observations made since 1997 at Parkes, we have detected all the
previously known pulsars and discovered nine new MSPs, 47~Tuc~O--W, all
of which are in binary systems.  Of these, 47~Tuc~R has the shortest
orbital period for any radio pulsar (96 minutes).  For details on the
search, orbital parameters, pulse profiles, and luminosities, see
Camilo et~al.~(2000).

\section{Pulsar timing}

The frequent detections of 14 pulsars allowed the determination of 12
new coherent timing solutions, and the confirmation of the two
previously known.  All pulsars lie in a circle of 1.5 arcmin about the
centre of the cluster (see Fig.~1).  Nine pulsars out of 14 have
negative period derivatives (see Table~1), indicating that they are
accelerating towards the Earth in the cluster's gravitational
potential, which can thus be probed in some detail.  The number of
pulsars with projected distance from the centre of the cluster smaller
than $R_{\perp}$ is proportional to $R_{\perp}$.  This is typical of an
isothermal distribution.  According to Phinney (1992) this indicates
neutron stars are the dominating mass species in the core of 47~Tuc.

\begin{figure}
\plotfiddle{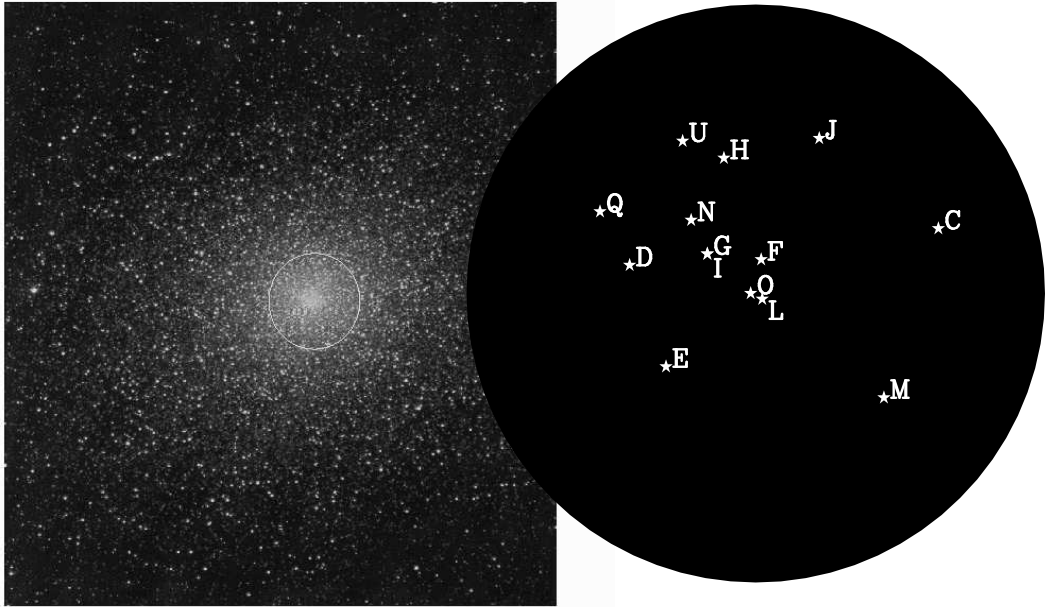}{3cm}{0}{40}{40}{-205}{-185}
\plotfiddle{pgplot.ps}{4cm}{270}{50}{50}{-120}{265}
\caption{Positions of 14 pulsars near the core of 47~Tuc.  The circle
has a radius of 100 arcseconds.}
\end{figure}

\end{document}